# Global Maps of Science based on the new Web-of-Science Categories

*Scientometrics* (in press)

Loet Leydesdorff,[1] Stephen Carley,[2] and Ismael Rafols[3]

In August 2011, Thomson Reuters launched version 5 of the *Science* and *Social Science Citation Index* in the Web of Science (WoS). Among other things, the 222 ISI Subject Categories (SCs) for these two databases in version 4 of WoS were renamed and extended to 225 WoS Categories (WCs). A new set of 151 Subject Areas was added, but at a higher level of aggregation. Perhaps confusingly, these Subject Areas are now abbreviated "SC" in the download, whereas "WC" is used for WoS Categories. Since we previously used the ISI SCs as the baseline for a global map in Pajek[4] (Rafols *et al*., 2010) and brought this facility online (at http://www.leydesdorff.net/overlaytoolkit), we recalibrated this map for the new WC categories using the *Journal Citation Reports* 2010. In the new installation, the base maps can also be made using VOSviewer[5] (Van Eck & Waltman, 2010).

The WC categories are sometimes different from the old SCs. For example, the *European Journal of Pharmaceutics and Biopharmaceutics*, which was previously classified to the ISI SC "Engineering, Chemical" was now assigned to "Pharmacology & Pharmacy." In an Annex, Rafols *et al.* (2010) showed that with aggregated sets the map is relatively robust against error in the classifications, but the differences can be significant at lower levels of aggregation (Rafols & Leydesdorff, 2009). As noted, the semantics may be confusing: WoS Subject Categories (WC) replace the old ISI Subject Categories (SC), but the old label SC for Subject Categories is now used in the download for the 151 newly created "Subject Areas" as they are called at the WoS interface. Both in the old and the new situation there are additionally 25 Subject Categories which are used exclusively in the Arts & Humanities Citation Index (A&HCI), and not included in our analysis because A&HCI does not have (hitherto) a *Journal Citation Report* (Leydesdorff, Hammarfelt & Salah, 2010, at pp. 2417 ff.). Table 1 summarizes the changes.

---
[1] Amsterdam School of Communication Research (ASCoR), University of Amsterdam, Kloveniersburgwal 48, 1012 CX Amsterdam, The Netherlands; loet@leydesdorff.net, http://www.leydesdorff.net.
[2] School of Public Policy, Georgia Tech, Atlanta, GA, USA; stephen.carley@gmail.com.
[3] SPRU (Science and Technology Policy Research), University of Sussex, Freeman Centre, Falmer Brighton, East Sussex BN1 9QE, United Kingdom; i.rafols@sussex.ac.uk.

[4] Pajek is freely available at http://vlado.fmf.uni-lj.si/pub/networks/pajek/ .
[5] VOSviewer is freely available at http://www.VOSviewer.com/ .

|  | WoS version 4 | WoS version 5 |
| --- | --- | --- |
| **Subject categories (SCI + SoSCI)** | 222 ISI Subject Categories[a] | 225 WoS Subject Categories |
| **Abbreviation in the download** | SC | WC |
| **A&HCI** | 25 ISI Subject Categories | 26 WoS Subject Categories |
| **Subject Areas** | <not defined> | 151 Subject Areas; abbreviated SC in the download |

[a] Three additional categories were no longer in use (Rafols & Leydesdorff, 2009).
[b] One category is currently not in use.[6]

**Table 1**: Changes in the organization of Subject Categories and Subject Areas between versions 4 and 5 of WoS.

Whereas previously 18 factors were found most appropriate for explaining the structure of the aggregated citation matrix, an organization into 19 factors explaining 54.3% of the variance is most apt for showing the disciplinary structure of the new matrix. The changes are mainly in the organization of mathematics. Rafols et al. (2010, at p. 1876) already noted that "the position of mathematics (all math subject categories) in the map remains open to debate. Since different strands of mathematics are linked to different major fields (medicine, engineering, social sciences), these may show as diverse entities in distant positions, rather than as a unitary corpus, depending on metrics, classifications, and clustering algorithms used."

In the new overlays, the previous factor of "Computer Science" is divided into two groups and designated (by us) as "Computer Science" and "Mathematical Methods." Table 2 provides the subject categories organized within these two compartments. The other 17 factors could remain similar to the previous classification (Rafols *et al.*, 2010).

| 12 WC attributed to "Computer Science" | 6 WCs attributed to "Mathematical methods" |
| --- | --- |
| Computer Science, Hardware & Architecture | Computer Science, Interdisciplinary Applications |
| Engineering, Electrical & Electronic | Operations Research & Management Science |
| Computer Science, Artificial Intelligence | Mathematics, Applied |
| Computer Science, Theory & Methods | Statistics & Probability |
| Computer Science, Information Systems | Engineering, Industrial |
| Telecommunications | Mathematics |
| Automation & Control Systems | |
| Computer Science, Cybernetics | |
| Computer Science, Software Engineering | |
| Robotics | |
| Imaging Science & Photographic Technology | |
| Transportation Science & Technology | |

**Table 2**: Distinction between "Computer Science" and "Mathematical Methods" in terms of WoS Subject Categories.

---

[6] One of the 225 WC is no longer in use; this is: "Biology, Miscellaneous". Only a single journal (*Growth Development and Aging*) is attributed to this category. Although this journal was cited 520 times in 2010 and is listed with an IF-2010 = 3.000, it is not processed in the *Journal Citation Reports* (JCR) from the citing side. In summary, we use the matrix of 224 WCs.

As with our previous installation, the base map with the relevant initialization routine for Pajek (v. 2.05) is available on the internet. The new base map can also be used as input to VOSviewer (Van Eck & Waltman, 2010), but this did not improve the visualization in the case of 19 clusters. (See for more discussion about the differences between the two visualization techniques Leydesdorff & Rafols (2012).) However, the clustering algorithm in VOSviewer distinguished four, in our opinion highly meaningful, groups (Waltman *et al.*, 2010). The organization into these four groups (biomedical, physical, environmental, and social sciences) is now provided as another (optional) partition within the Pajek (.paj) file containing the base map, and is also made available as input to VOSviewer.

The routine allows the user to make an overlay choosing one of the following two routes:

1. Within the WoS (v. 5.5), one can "Analyze Results" by clicking at the right top of the page with search results; choose "Web of Science Categories" among the ranking options, and export the data into a file "analyze.txt". This file can be read by the program wc10.exe, and the resulting file "wc10.vec" can be imported into the base map as a vector.[7] The visualization (*Draw-Partition-Vector* within Pajek) then shows the overlay (e.g., Figure 1).

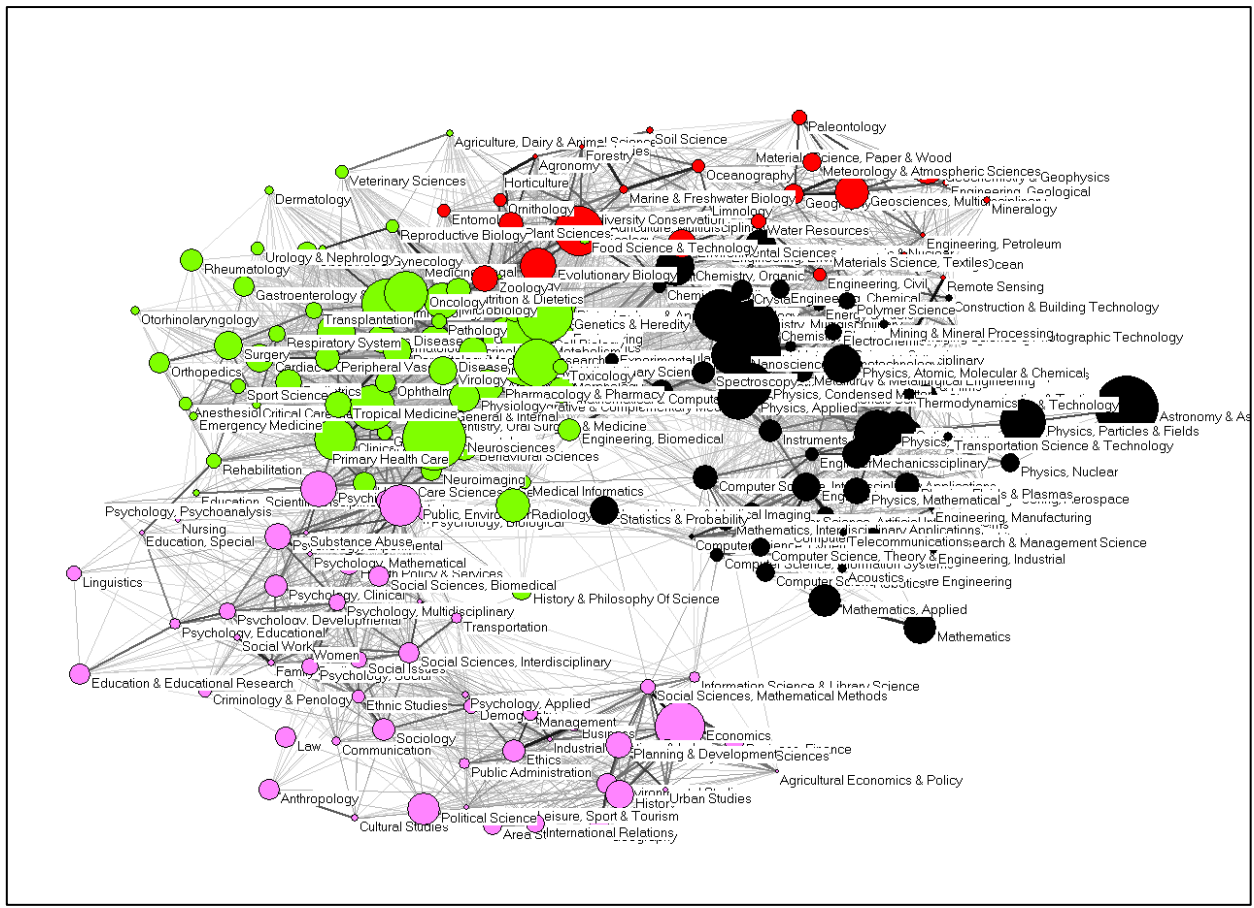

---

[7] The same file "wc10.vec" is automatically generated when using the routine ISI.exe for organizing the saved output of the WoS, as specified at http://www.leydesdorff.net/software/isi .

**Figure 1**: Disciplinary composition of 5,793 citable items (articles, proceedings papers, and reviews) published in 2009 with an address at Oxford University; Rao-Stirling = 0.856. (Pajek used for the visualization.)

2. The routine wc10.exe also generates three so-called "map"-files for VOSviewer: vos4.csv, vos6.csv, and vos19.csv. (The csv-extension makes these text files also readable using Excel.) These files generate maps when read into VOSviewer with four, six, and 19 clusters, respectively (using different colors). The manual of VOSviewer can be consulted for further options (Figure 2).

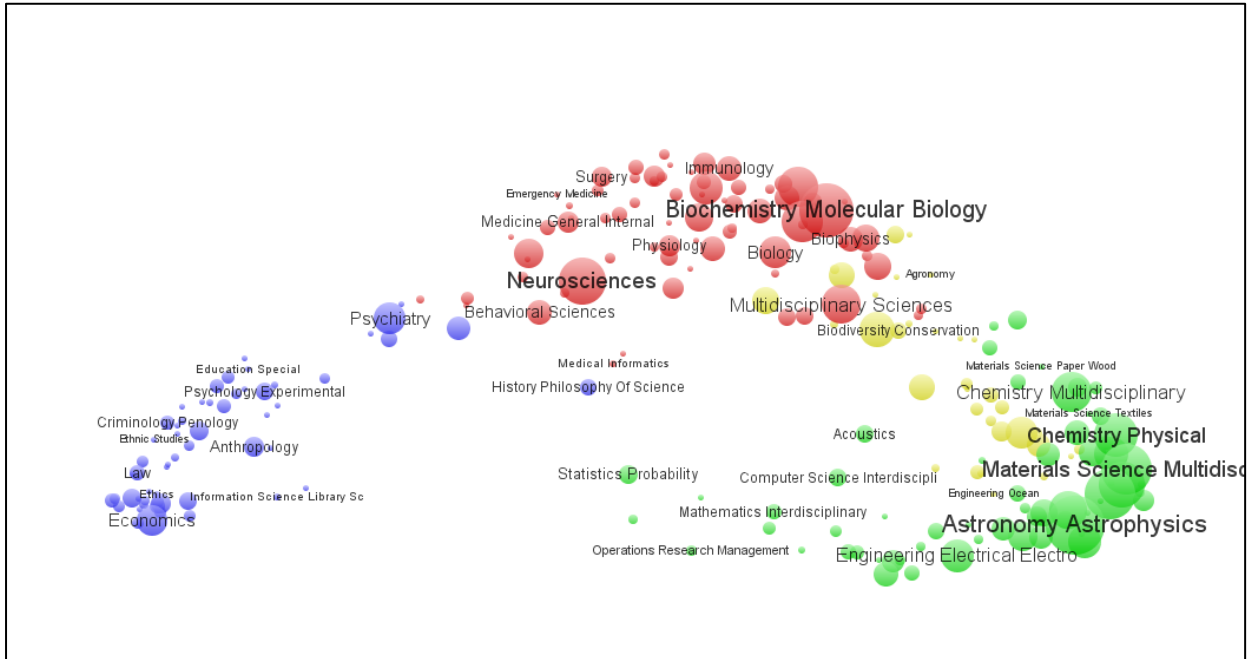

**Figure 2**: Disciplinary composition of 5,468 citable items (articles, proceedings papers, and reviews) published in 2009 with an address at the University of Cambridge; Rao-Stirling = 0.851. (VOSviewer used for the visualization.)

A PowerPoint file is provided at http://www.leydesdorff.net/overlaytoolkit/basemaps.pptx which allows the user to place the legends on top of the maps and which provides a few examples; these legends can also be edited. An additional routine enables the user to measure Rao-Stirling diversity as an index of the interdisciplinarity in the samples under study (cf. Carley & Porter, 2012; Leydesdorff & Rafols, 2011; Porter & Rafols, 2009; Rafols *et al*., in press; Stirling, 2007).

By thus adjusting to the new situation, we hope to have provided the community with a means to map inter- and multidisciplinary sets of documents in future research using the new version of the Web of Science (cf. Melkers and Hicks, 2012; Porter & Youtie, 2011; Soós & Kampis, 2011). A detailed manual for the mapping was provided (at the website) by Ken Riopelle. Additionally, a macro to transform the file "analyze.txt" into the Gephi format was provided by Clement Levallois.

**Acknowlegement**
We are grateful to Alan Porter for comments on a previous draft.
We acknowledge Thomson Reuters for the use of the data.